\documentclass[12pt,preprint]{aastex}
\slugcomment{Accepted for publication in ApJ}

\shorttitle{B/C measurement}
\shortauthors{Obermeier et al.}

\begin{document}
\title{The boron-to-carbon abundance ratio and Galactic propagation of cosmic radiation}

\author{A. Obermeier\altaffilmark{1,2}, P. Boyle\altaffilmark{1,3}, J. H\"orandel\altaffilmark{2}, and D. M\"uller\altaffilmark{1}}

\email{a.obermeier@astro.ru.nl}

\altaffiltext{1}{Enrico Fermi Institute, University of Chicago, Chicago, IL 60637, USA}
\altaffiltext{2}{Department of Astrophysics, Radboud Universiteit Nijmegen, 6525 HP Nijmegen, The Netherlands}
\altaffiltext{3}{Now at McGill University, Montreal, Canada}

\begin{abstract}
In two long-duration balloon flights in 2003 and 2006, the TRACER cosmic-ray detector has measured the energy spectra and the absolute intensities of the cosmic-ray nuclei from boron ($Z=5$) to iron ($Z=26$) up to very high energies. In particular, the second flight has led to results on the energy spectrum of the secondary boron nuclei, and on the boron abundance relative to that of the heavier primary parent nuclei, commonly quantified as the ``B/C abundance ratio''. The energy dependence of this ratio, now available up to about 2 TeV per amu, provides a measure for the energy dependence of cosmic-ray propagation through the Galaxy, and for the shape of the cosmic-ray source energy spectrum. We use a Leaky-Box approximation of cosmic-ray propagation to obtain constraints on the relevant parameters on the basis of the results of TRACER and of other measurements. This analysis suggests that the source energy spectrum is a relatively soft power law in energy $E^{-\alpha}$, with spectral exponent $\alpha=2.37\pm0.12$, and that the propagation path length $\Lambda(E)$ is described by a power law in energy with exponent $\delta=0.53\pm0.06$, but may assume a constant residual value $\Lambda_0$ at high energy. The value of $\Lambda_0$ is not well constrained but should be less than about 0.8~g~cm$^{-2}$. Finally, we compare the data with numerical solutions of a diffusive reacceleration model, which also indicates a soft source spectrum.

\end{abstract}


\section{Introduction}
\label{sec:intro}

The composition and energy spectra of Galactic cosmic rays change while the particles propagate from the acceleration site to the observer. Thus, the cosmic-ray population observed in measurements is different from the source population, even in the approximation that the ambient population is the same in time and space throughout the volume of the Galaxy. To determine the characteristics of the cosmic-ray sources, the propagation effects must be deconvoluted from the measured data.

Processes affecting the cosmic rays after they are released from their sources include diffusion and diffusive or convective escape from the Galaxy; interactions with the components of the interstellar gas leading to the loss of the primary and to the production of secondary particles; decay of radio-active components; secondary acceleration in the interstellar medium (ISM); ionization energy loss; or radiative energy loss (for electrons). Before the cosmic rays reach an observer near the Earth, they may also be affected by solar modulation. Many of these processes (including their energy dependence) are more difficult to quantify at relatively low energies, below a few GeV per amu, than in the highly relativistic region. Nevertheless, even at high energies, the number of parameters that appear in the full transport equation for Galactic cosmic rays (see e.g.\ \citet{ginzburg}) is too large, and the parameters are often too poorly known, as to permit an easy analytic description of the effects of cosmic-ray propagation. In the following, we will review recent experimental data in the context of the long-popular ``Leaky-Box'' approximation of particle transport in the Galaxy. We also will compare the data with a diffusive propagation model using the prescriptions of the GALPROP code~\citep{galprop,GPwebrun}.

At low energies, observational data of good precision have been available for some time from detailed measurements with a number of spacecraft. For the high energy region, from GeV~amu$^{-1}$ to TeV~amu$^{-1}$ energies\footnote{The measured quantity for the TRACER instrument (and many others) is the Lorentz-factor $\gamma=E/mc^2$, which is equivalent to (kinetic energy per amu + 1) if the mass is expressed in amu. Within the uncertainties of most measurements, the kinetic energy per amu is identical to the more commonly used quantity kinetic energy per nucleon.}, few measurements in space are available, but a number of long-duration balloon (LDB) flights in recent years have provided new data. The following study will therefore concentrate on the high-energy region.

Specifically, this study is motivated by recent results from the TRACER cosmic-ray detector. This instrument is currently the largest and most sensitive detector for the heavier ($Z>3$) cosmic-ray nuclei~\citep{AveNim}. TRACER was flown in two LDB flights, in 2003 in Antarctica (LDB1), and in 2006 from Kiruna, Sweden (LDB2). The 2003 flight yielded a comprehensive set of the spectra of the major primary nuclei from oxygen ($Z=8$) to iron ($Z=26$), covering the energy range from a few GeV per amu to several TeV per amu~\citep{AveMeas}. For the 2006 flight of TRACER, several significant upgrades of the instrument permitted the inclusion of the light elements of boron ($Z=5$), carbon ($Z=6$) and nitrogen ($Z=7$) in the measurement~\citep{OberMeas}. Particularly important are the measurements of boron nuclei which are purely of secondary interstellar origin. Together with the earlier measurements in space with HEAO-3~\citep{HEAObc}, CRN~\citep{CRNbc,crn}, and with AMS-01~\citep{ams}, and with the results from the LDB flights of ATIC~\citep{ATICspectra,ATICbc}, and CREAM~\citep{CREAMbc,CREAMspectra}, the current data set on the composition of high-energy cosmic-ray nuclei should lead to more stringent constraints on the propagation of high-energy cosmic rays than previously possible.

\section{Parameters of Cosmic Ray Propagation}

The ``Leaky-Box'' approximation parametrizes the diffusion of cosmic rays in the Galaxy by introducing the average propagation path length $\Lambda(E)$, which is inversely proportional to the diffusion coefficient and may depend on energy $E$:
\begin{equation}
  \Lambda(E) = \beta c\rho\tau(E),
\label{eq:Lesc_tau}
\end{equation}
where $\tau(E)$ is the average containment time in the Galaxy, $\beta=v/c$ is the particle's velocity, and $\rho$ is the mass density of the interstellar gas. Similarly, the average spallation path length $\Lambda_s(A)$ quantifies the interaction of a cosmic-ray nucleus with mass number $A$ by spallation reactions with interstellar gas nuclei of mass $M$:
\begin{equation}
 \Lambda_s(A) = \frac{M}{\sigma(A)}.
\label{eq:Ls_sigma}
\end{equation}
For relativistic energies, we assume that the spallation cross section $\sigma$ depends only on $A$, ignoring a possible small energy dependence. The values of $\Lambda$  and $\Lambda_s$ are of the same order of magnitude at GeV energies.

For highly relativistic nuclei, one may disregard ionization energy losses and solar modulation. If one also ignores convective motion of particles as well as secondary acceleration and radio-active decay, one obtains a continuity equation for the ambient differential density $N_i$ of cosmic-ray species $i$ in the form~(e.g.~\citet{AveInt}):
\begin{equation}
 N_i(E)=\frac{1}{\Lambda(E)^{-1}+\Lambda_s(A)^{-1}}\left(\frac{Q_i(E)}{\beta c \rho}+\sum_{k>i}\frac{N_k}{\lambda_{k-i}}\right)
\label{eq:LB}
\end{equation}
Here, $Q_i(E)$ is the rate of production in the source, and $\lambda_{k-i}$ is the differential spallation path length for a nucleus $k$ to spallate into $i$. As is common practice, we will accept this form of the Leaky-Box equation as a valid approximation for the energy spectra of the more abundant cosmic-ray species.

The Leaky-Box approximation assumes a continuous distribution of cosmic-ray sources and makes no assumption about the boundaries of the diffusion region and the extent of the Galactic halo. Nevertheless, it has been found (e.g. \citet{ptuskin1}) that  more realistic diffusion models, such as the numerical integration of the transport equation in the GALPROP code \citep{galprop}, lead to results for the major stable cosmic-ray nuclei, which are equivalent to the Leaky-Box predictions at high energy.

The propagation path length $\Lambda$ decreases with energy above a few GeV~amu$^{-1}$. The first evidence for this phenomenon came from the observation of the decreasing relative abundance of secondary, spallation-produced cosmic-ray nuclei with energy~\citep{Juliusson, smith}. This and subsequent measurements, in particular on HEAO-3~\citep{HEAObc} and CRN~\citep{CRNbc}, indicated that the energy dependence of $\Lambda$ might have the form of a power law $E^{-\delta}$, with $\delta\approx0.6$, at relativistic energies ($>20$~GeV~amu$^{-1}$).

It is of historic interest that at about the same time, stochastic shock-acceleration in supernova-remnants was proposed as an efficient process to produce Galactic cosmic rays (e.g.~\citet{Bell1978}). However, for strong shocks, this theory predicts in, to first order, a source energy spectrum $Q_i(E)$ in form of a power law $E^{-\alpha}$, with $\alpha\approx2$, much harder than the well-known $E^{-2.7}$ behavior of the observed spectrum. A cosmic-ray escape from the Galaxy that scales with $E^{-0.6}$ indeed seemed to provide an easy solution to this dilemma. This fact may well have fostered the acceptance of the shock acceleration model. However, a closer look at currently available data as will be attempted in this paper, may reveal complications to this simple picture.

The $E^{-\delta}$ energy dependence of the propagation path length would lead to very small values of $\Lambda$ at high energies that might not be consistent with the reported isotropy of the cosmic-ray flux. Therefore, one may speculate that the path length approaches a residual value $\Lambda_0$ at high energies:
\begin{equation}
 \Lambda=C\cdot E^{-\delta}+\Lambda_0.
\label{eq:Lesc_L0}
\end{equation}
Physically, the residual path length would characterize a minimum column density of matter that a cosmic-ray particle must traverse, even at very high energy. This matter could be located near the cosmic-ray source, or it could signify the minimum distance to a source.

It is commonly assumed, and consistent, to first order, with the shock acceleration hypothesis, that the energy spectrum of all primary cosmic-ray nuclei, has the same dependence on energy (or rigidity), typically in form of a power law with common index $\alpha$:
\begin{equation}
 Q_i(E)=n_i\cdot E^{-\alpha},
\label{eq:source}
\end{equation}
Equation~(\ref{eq:LB}) then takes the form
\begin{equation}
 N_i(E)=\frac{1}{(C\cdot E^{-\delta}+\Lambda_0)^{-1}+\Lambda_s(A)^{-1}}\left(\frac{n_i\cdot E^{-\alpha}}{\beta c \rho}+\sum_{k>i}\frac{N_k}{\lambda_{k-i}}\right).
\label{eq:LB_2}
\end{equation}

This equation describes the connection at high energies between the measured cosmic-ray spectrum $N_i(E)$ for every stable nuclear species $i$, and the power-law spectrum and relative intensity $n_i$ of that species at the source. Besides a normalization factor $C$, this relation uses just three free parameters, the source index $\alpha$, the propagation index $\delta$, and the residual path length $\Lambda_0$.

The equation illustrates in a simple way how the competition between diffusive and spallation loss, expressed by the term $1/(\Lambda(E)^{-1} + \Lambda_s(A)^{-1})$, is reflected in the shape of the measured energy spectra: the smaller of the two $\Lambda$-parameters will most strongly affect the resulting spectrum. Current data, which will be further discussed below, and which are summarized in Figure~\ref{fig4}, show that $\Lambda(E)$ is smaller than $\Lambda_s$ for much of the energy region of concern. However, to completely ignore the influence of spallation on the spectral shape we must require that $\Lambda(E)$ is smaller than $\Lambda_s$ by at least a factor of ten. For the nuclei of concern here, this will be the case for energies in the 100 -- 1000~GeV~amu$^{-1}$ region if the residual path length $\Lambda_0$ is zero (as is illustrated in Figure~\ref{fig4}). At these energies, most measured data are affected by significant statistical errors and do not strongly constrain the functional shape of the observed spectrum. Returning to Equation~(\ref{eq:LB_2}), we predict that the observed spectrum will be softer than the source spectrum $E^{-\alpha}$, but that the often used description of the observed spectrum as a power law $E^{-\Gamma}$ with index $\Gamma=(\alpha + \delta)$ can only become valid at the highest energies, and if the residual path length is zero. At lower energies (and that is where the most accurate data are available), a power-law fit with a smaller value, $\Gamma<(\alpha + \delta)$ would approximate the observed spectrum. 

\section{The Data}
\label{sec:data}

\begin{figure}[tbh]
 \includegraphics[width=\linewidth]{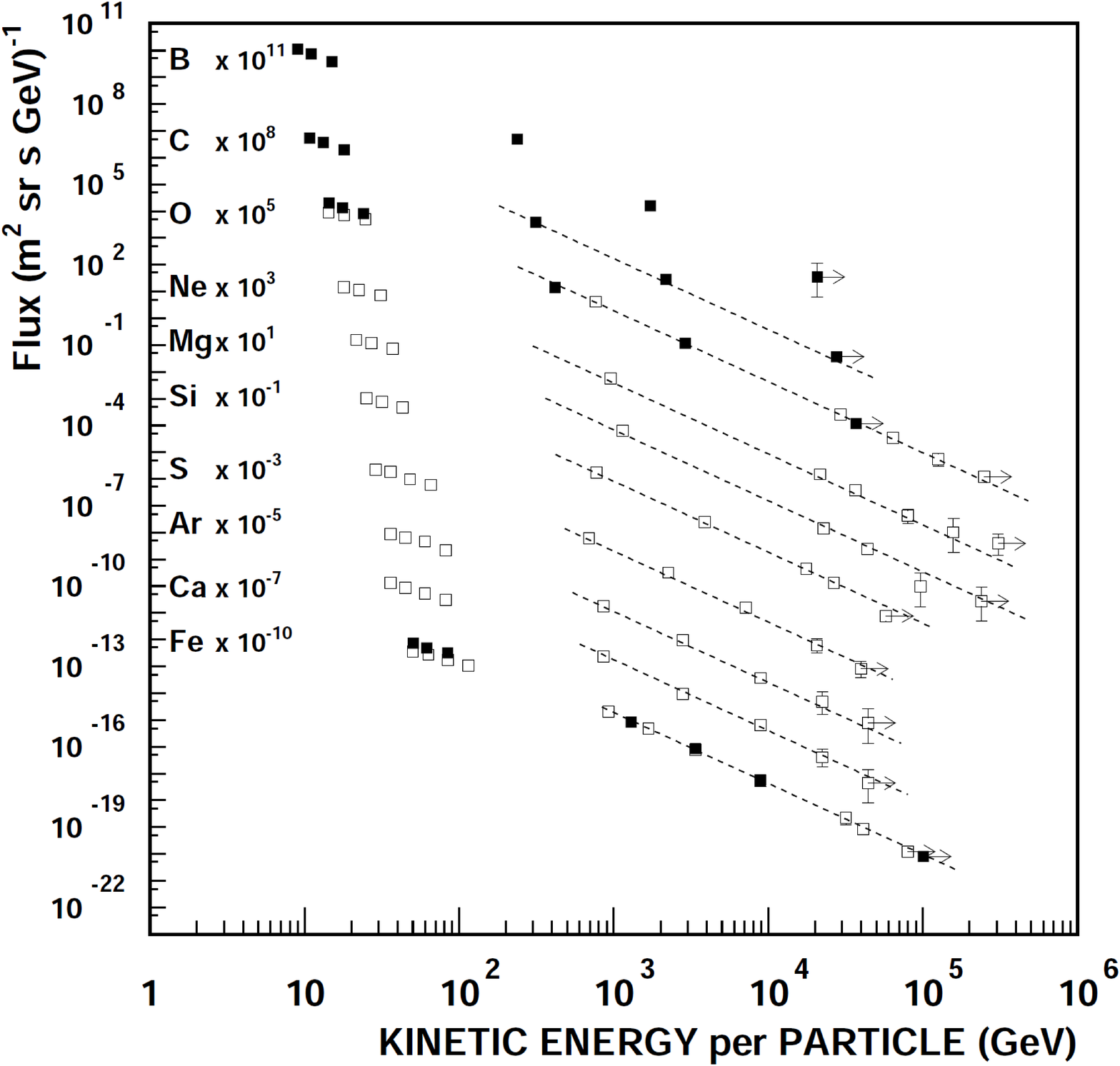}
 \caption{Compilation of the differential energy spectra measured by TRACER in LDB1 (open symbols) and LDB2 (solid symbols) (see \citet{AveMeas} and \citet{OberMeas}). Dashed lines indicate a power-law fit above 20~GeV~amu$^{-1}$.\label{fig1}}
\end{figure}

The measurements available from the two long-duration balloon flights of TRACER cover the energy spectra of ten cosmic-ray elements from boron ($Z=5$) to iron ($Z=26$). The data of the first flight, for the primary nuclei from oxygen ($Z=8$) to iron, have been reported by~\citet{AveMeas}, and the results of the second LDB flight are described by~\citet{OberMeas}.  Where overlap exists, the two data sets agree well with each other. The resulting energy spectra are shown in Figure~\ref{fig1}. Figure~\ref{fig1} also indicates a simple power law fit with index $2.65\pm0.05$ which was found to describe all primary spectra quite well above 20~GeV~amu$^{-1}$, without any significant change with charge number $Z$~\citep{AveInt}.

\begin{figure}[tbh]
 \includegraphics[width=\linewidth]{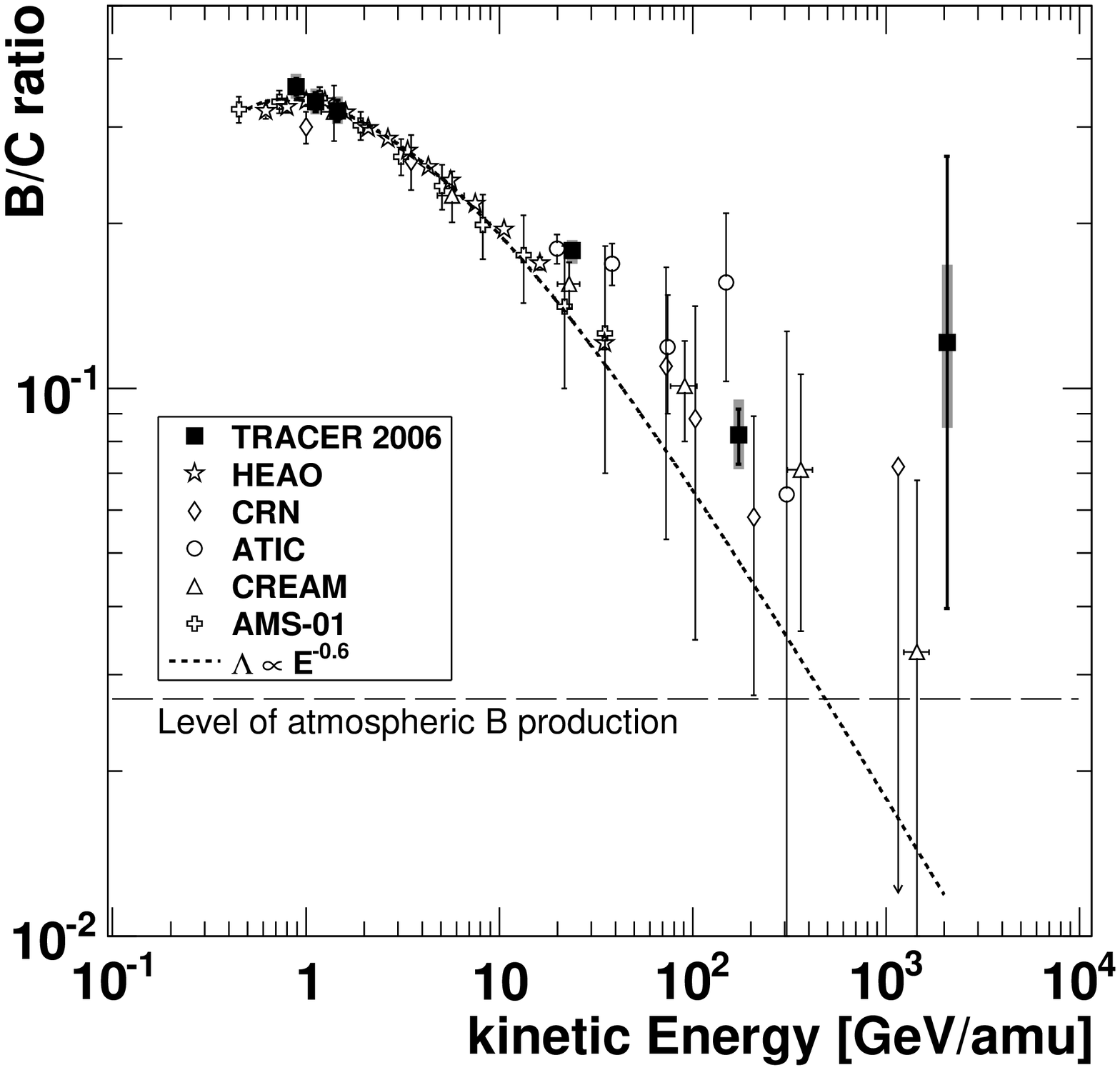}
 \caption{Boron-to-carbon abundance ratio as a function of kinetic energy per nucleon as measured by TRACER~\citep{OberMeas}, HEAO~\citep{HEAObc}, CRN~\citep{CRNbc}, ATIC~\citep{ATICbc}, CREAM~\citep{CREAMbc} and AMS-01~\citep{ams}. Error bars are statistical (thin) and systematic (thick, only for TRACER). A simple model of the escape pathlength is indicated (dotted, see Eq.~\ref{eq:yanasak}). For the TRACER measurement the level of the subtracted contribution of atmospheric production of boron (dashed) is shown.\label{fig2}}
\end{figure}

The data of the second flight include the energy spectra for the light elements boron ($Z=5$) and carbon ($Z=6$) up to about 2~TeV~amu$^{-1}$. The resulting boron-to-carbon (B/C) abundance ratio~\citep{OberMeas} is shown in Figure~\ref{fig2}. This figure also includes previous data from HEAO-3~\citep{HEAObc}, CRN~\citep{CRNbc}, ATIC~\citep{ATICbc}, CREAM~\citep{CREAMbc}, and AMS-01~\citep{ams}, with their reported statistical uncertainties. At the highest energies, all results are based on few events, and not all measurements then define the statistical uncertainty in the same way. The results on the B/C ratio from balloon flights include a correction for atmospheric production of boron which may become sizable at high energy. The level of the correction for the TRACER data, which has been subtracted from the ratio (at an average residual atmosphere of 5.2~g~cm$^{-2}$), is indicated as a dashed line in Figure~\ref{fig2}~\citep{AtmoSec}. Details about this correction will be published separately.

Also shown in Figure~\ref{fig2}, is a prediction for the B/C ratio corresponding to an energy dependence of the escape path length $\Lambda$, essentially proportional to $E^{-0.6}$.  Specifically, we have chosen for this prediction the parametrization of~\citet{yanasak}, which was developed to include low-energy data from HEAO-3 and ACE-CRIS (below $\sim10$~GeV~amu$^{-1}$):
 \begin{equation}
  \Lambda(R)\;[\rm{g~cm}^{-2}]=\frac{C\beta}{(\beta R)^{\delta} + (0.714\cdot \beta R)^{-1.4}}+\Lambda_0,
 \label{eq:yanasak}
 \end{equation}
where $R=pc/Ze$ is the particle's rigidity and $\beta=v/c$ is the particle's velocity. \citet{yanasak} use a propagation index $\delta=0.58$, a residual path length $\Lambda_0=0$, and a normalization $C=26.7$. It should be noted that at high energies Equation~(\ref{eq:yanasak}) is equivalent to Equation~(\ref{eq:Lesc_L0}).

\section{Constraints on Propagation Parameters}
\label{sec:LB}

We now will attempt to derive constraints on the propagation parameters from the measurements just described, i.e.\ constraints on the cosmic-ray source spectral index $\alpha$, and on the energy-dependence of the propagation path length characterized by the parameters $\delta$ and $\Lambda_0$. The present work continues the fitting procedures by~\citet{AveInt} applied to the data of the first LDB flight of TRACER in 2003.

Referring the reader to the paper by~\citet{AveInt}, we recall that this first flight led to the energy spectra of the heavier primary cosmic-ray nuclei, but did not include results on secondary elements such as boron~\citep{AveMeas}. Hence, no new information on the propagation path length was obtained, and for the fitting procedure fixed parameters $\delta=0.6$ and $C=26.7$ were assumed. However, $\Lambda_0$ and $\alpha$ were treated as free parameters. The most striking feature of the measured data was the common power-law appearance of all measured energy spectra from 20~GeV~amu$^{-1}$ to several TeV~amu$^{-1}$ (with the same index of 2.65). This feature could only be reconciled with the prediction of Equation~(\ref{eq:LB_2}), if the energy spectrum at the source was fairly soft, with a probable value of the source index $\alpha$ between 2.3 and 2.45. This value is considerably larger than the first-order expectation of $\alpha\approx2.0$ for acceleration in strong shocks. The data from this flight did not place strong constraints on the residual path length $\Lambda_0$, and could not exclude a non-zero value for this parameter. With these fitting results, the measured energy spectra of the individual elements were extrapolated back to the sources, and the relative elemental source abundances $n_i$ had been obtained~\citep{AveInt}.

With the measurement of the energy spectrum of the secondary nucleus boron, and of the secondary/primary intensity ratio, i.e.\ the B/C ratio, in the second balloon flight in 2006, we now attempt to derive further detail. We use Equation~(\ref{eq:LB}), which for boron does not contain a source term $Q_i$. Introducing an effective path length $\lambda_{\rightarrow B}$ (see Equation~(\ref{eq:Leff})), the B/C ratio can then be expressed as:
\begin{equation}
 \frac{N_B}{N_C}=\frac{\lambda_{\rightarrow B}^{-1}}{\Lambda^{-1} + \Lambda_B^{-1}},
\label{eq:BC}
\end{equation}
Here, we further assume that boron is produced only by spallation of carbon and oxygen, i.e.\ the contributions from the spallation of nitrogen (amounting to just $\sim3$\% of the boron intensity) and from nuclei with $Z>8$ are ignored. Finally, we assume that there are no significant contributions to the intensities of carbon and oxygen from spallation of heavier nuclei. These assumptions seem to be justified by the dominant intensities of carbon and oxygen among the primary nuclei. The effective production path length for boron $\lambda_{\rightarrow B}$ includes both carbon and oxygen as parent nuclei:
\begin{equation}
 \lambda_{\rightarrow B}^{-1}  = \lambda_{C\rightarrow B}^{-1}  + N_O/N_C\cdot\lambda_{O\rightarrow B}^{-1}.
\label{eq:Leff}
\end{equation}
The ratio $N_O/N_C$ refers to the intensity ratio of the parent nuclei oxygen and carbon on top of the atmosphere. This ratio can be taken as independent of energy, and is close to unity~\citep{OberMeas,crn,HEAObc,CREAMbc}. The spallation path length $\Lambda_B$ in Equation~(\ref{eq:BC}) is derived from a geometrical parametrization of the cross sections~\citep{BPform,westfall}, and the production path lengths $\lambda$ in Equation~(\ref{eq:Leff}) are derived from partial cross sections determined by~\citet{webber2}. Specifically, we use $\Lambda_B=9.3$~g~cm$^{-2}$, and $\lambda_{\rightarrow B}=26.8$~g~cm$^{-2}$ (assuming the interstellar medium as a mixture of 90\% hydrogen and 10\% helium by number).

The fitting function is then given with Equation~(\ref{eq:BC}), with the escape path length $\Lambda$ as expressed in Equation~(\ref{eq:yanasak}). Compared to using the high-energy form of Equation~(\ref{eq:Lesc_L0}), this has the advantage that data below $\sim$10~GeV~amu$^{-1}$ can be included in the fit. The only unknown quantity in Equation~(\ref{eq:BC}) is the energy dependence of the propagation path length $\Lambda$ with the parameters $\delta$ and $\Lambda_0$.

\begin{figure}[tbh]
 \includegraphics[width=\linewidth]{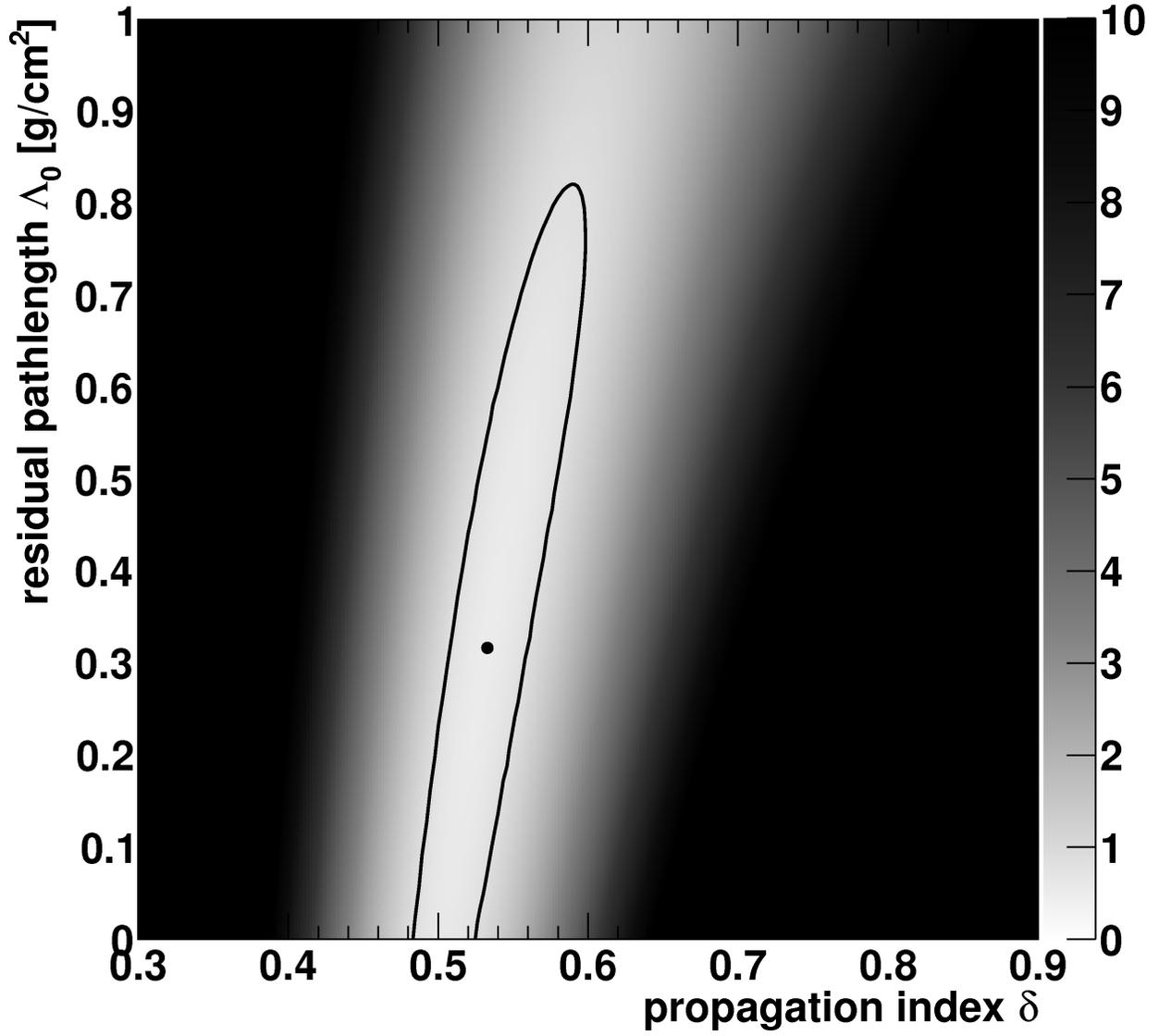}
 \caption{$\chi^2$ map in the parameter space of $\delta$ vs. $\Lambda_{0}$ for the Leaky-Box model fit to TRACER data. The best fit values are marked at $(\delta,\Lambda_0)=(0.53\pm0.06,0.31^{+0.55}_{-0.31}\,\rm{g~cm}^{-2})$ and the 1$\sigma$ contour is indicated. \label{fig3}}
\end{figure}

\begin{figure}[tbh]
 \includegraphics[width=\linewidth]{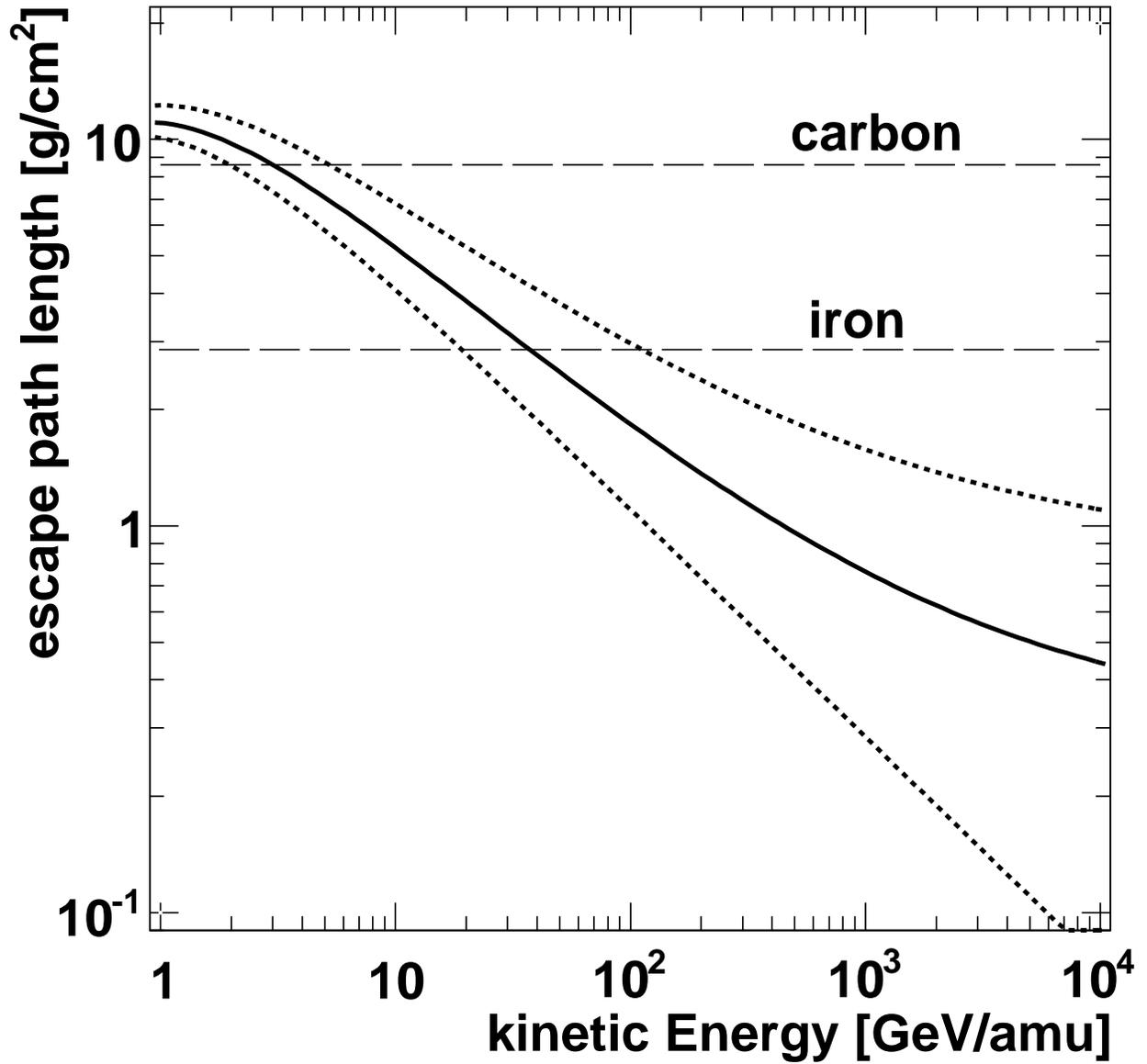}
 \caption{Escape path length as a function of energy resulting from a fit to the boron-to-carbon data of TRACER. The dotted lines indicate the uncertainty range noted in Fig.~\ref{fig3}. The dashed lines indicate the spallation path lengths of carbon and iron in the interstellar medium. \label{fig4}}
\end{figure}

\begin{figure}[tbh]
 \includegraphics[width=\linewidth]{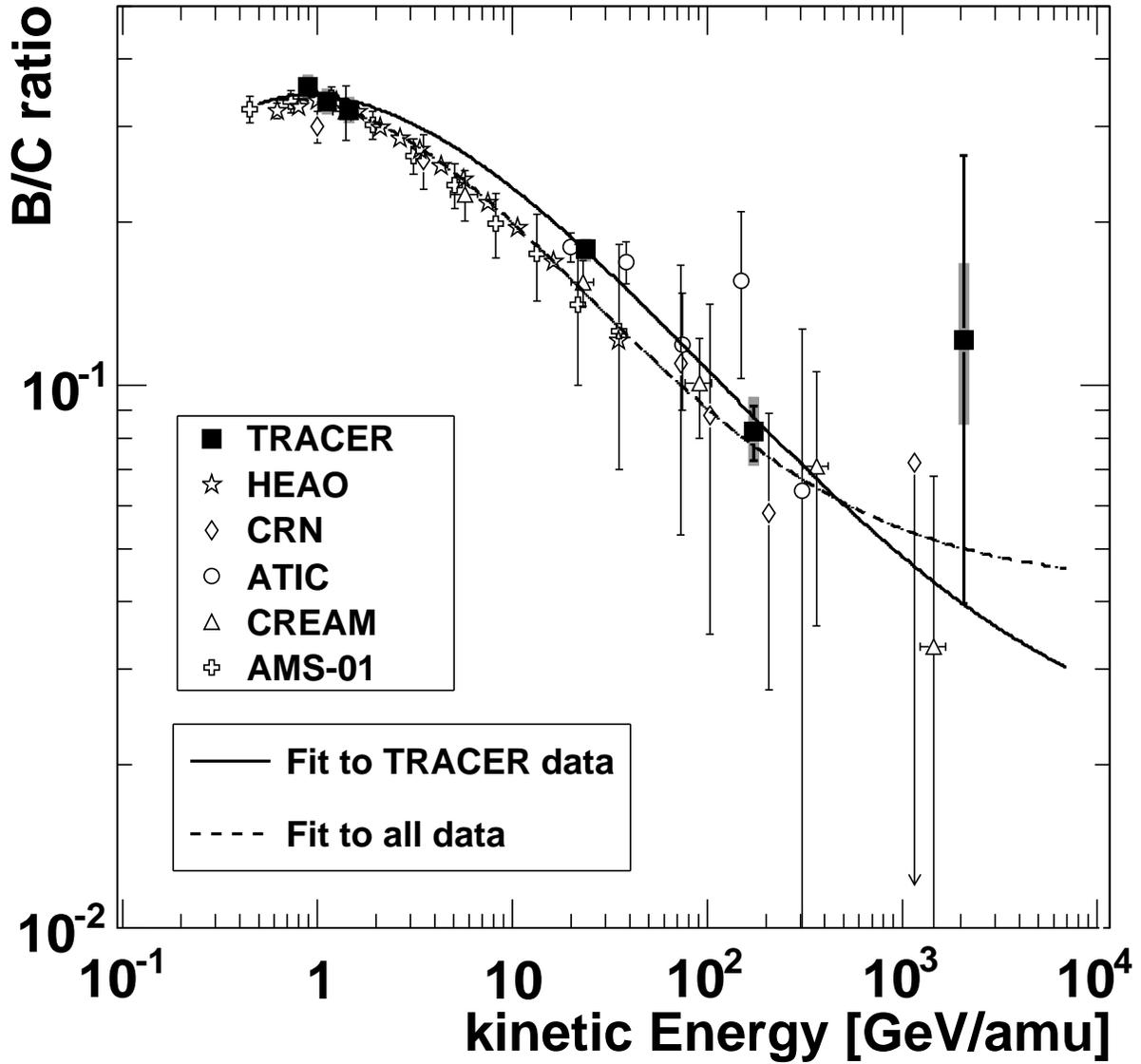}
 \caption{Abundance ratio of boron and carbon from TRACER~\citep{OberMeas} and from other measurements (see Fig.~\ref{fig2}). The solid line represents the best fit to the TRACER data alone; the dashed line is the best fit to all data combined. \label{fig5}}
\end{figure}

We have fitted the data on the B/C ratio versus energy as measured by TRACER to a variety of values for $\delta$ and $\Lambda_0$. A probability contour map of the fitting results is shown in Figure~\ref{fig3}. The best fit for the propagation index is $\delta = 0.53\pm0.06$~g~cm$^{-2}$, and is quite close to the value of 0.6 which was used in the previous analysis of~\citet{AveInt}. The best value for the residual path length,  $\Lambda_0 = 0.31^{+0.55}_{-0.31}$~g~cm$^{-2}$, is less well defined, and still a solution with $\Lambda_0 = 0$ cannot be excluded within the present accuracy of the TRACER data alone. The corresponding escape path length $\Lambda$ together with its uncertainties is shown in Figure~\ref{fig4} as a function of energy. The figure indicates that a cosmic-ray nucleus most probably traverses a column density of $2.5\pm0.9$~g~cm$^{-2}$ of matter at an energy of 50~GeV~amu$^{-1}$ before escaping the Galaxy. At 1000~GeV~amu$^{-1}$, the path length will be between 1.6~g~cm$^{-2}$ and 0.28~g~cm$^{-2}$, with a best-fit value of 0.76~g~cm$^{-2}$. For comparison, the figure also indicates the energy-independent spallation path lengths for the primary elements carbon and iron. The result of the fitting procedure is shown in Figure~\ref{fig5} as a solid line. The fit to the TRACER data alone overshoots the low energy data of other measurements by about 10\%--20\%.

To refine the fit we may attempt to use the total data set currently available for all reported B/C ratio measurements at high energy (see Figure~\ref{fig2}) in the fitting routine. The result for the propagation parameters of this analysis essentially agrees with the analysis of the TRACER data alone, but leads to values which are more tightly constrained: we now obtain $\delta = 0.64 \pm 0.02$, and $\Lambda_0 = 0.7 \pm 0.2$~g~cm$^{-2}$. If this is correct, it would be the first evidence for a non-zero residual path length. However, we feel that this conclusion must be taken with caution as the different measurements may be affected by different systematic and statistical uncertainties, including uncertainties in the cross sections and the contribution of nitrogen. The result of the fit to all data is also illustrated in Figure~\ref{fig5} (dashed line).

Turning our attention now toward the shape of the energy spectra of primary nuclei, we
show in Figure~\ref{fig6} the energy spectrum for oxygen, $N_O(E)$ (multiplied with
$E^{2.65}$), as measured in the two TRACER flights. The prediction of Equation~(\ref{eq:LB_2}) is applied with the propagation parameters from TRACER as just described ($\delta = 0.53 \pm 0.06$, $\Lambda_0 = 0.31^{+0.55}_{-0.31}$~g~cm$^{-2}$). When the source spectral index $\alpha$ is varied, a best fit for the source spectrum is obtained with $\alpha=2.37\pm0.12$. The shape of this source spectrum is indicated in Figure~\ref{fig6}. As mentioned, the production of oxygen by spallation is ignored in the fitting procedure, and the spallation path length of oxygen is derived from a geometrical parametrization~\citep{BPform,westfall}. As in our earlier analysis~\citep{AveInt}, this fit again indicates a fairly soft energy spectrum at the source. One may note that the corresponding energy spectrum expected to be observed at the Earth is not a straight power law. The curvature reflects the effect of the non-zero residual path length. A second deviation from a straight power-law behavior of the observed spectrum is expected at lower energy due to the competition of spallation and escape during galactic propagation, and eventually due to solar modulation. Unfortunately, the uncertainties in the data are too large as to determine whether these subtle curvature effects can be real.

\section{Comparison with a diffusive propagation model}
\label{sec:GALPROP}

A more realistic approach to describe the cosmic-ray propagation in the Galaxy involves numerical solutions of the transport equations. The most commonly used tool is the GALPROP code~\citep{galprop,GPwebrun}. It has been successfully employed to provide a self-consistent description of most observations of cosmic rays including protons and anti-protons, heavier nuclei, electrons (but apparently not positrons), over a wide range of energies, and of diffuse Galactic gamma rays. 

We apply GALPROP with input parameters (e.g. scale height, cross sections, source abundances) as suggested in several publications~\citep{GP1,GP2}. We also accept a diffusive propagation model with an energy-dependent diffusion coefficient commensurate with Kolmogorov turbulence in the Galactic magnetic field, i.e., $D$ proportional to $E^\delta$, with $\delta\approx1/3$. The model also permits reacceleration of cosmic rays in interstellar space. This model is widely used to describe cosmic radiation in the Galaxy and the values of its parameters have been applied for some time~\citep{ptuskin2,GP1} and have recently been confirmed again by~\citet{trotta}.

In this model, no asymptotic value for the diffusion coefficient is invoked, so there is no equivalent to the residual path length $\Lambda_0$. For this specific model, the preferred value for the power-law index of the cosmic-ray source spectra is $\alpha = 2.34$, very similar to the value resulting from our Leaky-Box fit described above.

We compare the prediction of this model with the measurements of the B/C ratio, as shown in Figure~\ref{fig7}. Indeed, the model describes the data well, and closely constrains the value of the propagation parameter to $\delta\approx0.34$. This value is smaller than the parameter $\delta = 0.53\pm0.06$ obtained for the Leaky-Box fit.

The choice of parameters in the diffusion model is not unique. While it is not the purpose of this study to review the various parameter selections that have been discussed in the literature, we emphasize that reacceleration affects the cosmic-ray energy spectra in the diffusive model, and contributes to the energy dependence of the B/C ratio. If reacceleration were insignificant, the diffusion model would be commensurate with a value $\delta\approx0.6$ obtained in the Leaky Box approximation, but the source spectral index $\alpha$ would be smaller than before, $\alpha\approx2.15$ (e.g.~\citet{ptuskin2}).

At the very highest energies, a non-zero value of the residual path length $\Lambda_0$ would flatten the energy dependence of the B/C ratio. While this would be a distinct signature not predicted in the diffusion model (unless a primary contribution to boron is invoked), its observation is still within the fringes of current experimental accuracy.

\section{Discussion and Conclusion}

We have discussed constraints on the propagation of cosmic rays through the Galaxy derived from the measurements with the TRACER instrument, and from several other recent observations. A simple Leaky-Box model is used that reduces the number of free parameters to the spectral index $\alpha$ of the source energy spectrum (assumed to have the same power-law form for all primary nuclei), and to the energy dependence of the diffusive escape path length $\Lambda_{\rm{esc}}$, which decreases with energy as a power law with index $\delta$, but may also exhibit a constant residual value $\Lambda_0$ (see Equation~(\ref{eq:Lesc_L0})).

A previous analysis had been performed based on the data on the primary cosmic-ray nuclei from oxygen ($Z=8$) to iron ($Z=26$), obtained with the first LDB flight of TRACER~\citep{AveInt}. This analysis had led to the conclusion that the energy spectra of the cosmic rays at the sources must be quite soft, with $\alpha=2.3-2.4$. No new data on secondary cosmic rays were available at that time; hence, a propagation index $\delta=0.6$ was assumed in the analysis, and no strong constraint on $\Lambda_0$ could be obtained.

The data discussed in the present work include results from the second LDB flight of TRACER~\citep{OberMeas}, which provides a new measurement of boron and carbon nuclei, and of the B/C ratio at high energies. Therefore, the fitting routine could include the parameters, $\alpha$, $\delta$, and $\Lambda_0$ as free parameters. Remarkably, the best-fit values of these parameters agree well with the conclusions of the previous work: again, the source spectrum is quite soft, with $\alpha=2.37\pm0.12$ and the propagation index $\delta=0.53\pm0.06$ is close to the previously fixed value of 0.6. The residual path length $\Lambda_0$ may be about 0.3~g~cm$^{-2}$ but still has a large uncertainty which cannot exclude a value $\Lambda_0=0$.

The agreement with the earlier work has an important consequence: the relative abundances of the individual elements at the sources (i.e., the numbers $n_i$ in Equations~(\ref{eq:source}) and (\ref{eq:LB_2})), which had been calculated in the earlier analysis of~\citet{AveInt}, remain valid. In particular, the comparison of these values with the solar system abundances (often also called universal abundance scale~\citep{grevesse,lodders}), and the correlations with parameters such as the first ionization potential, or the condensation temperature, do not need to be updated.

The softness of the energy spectra at the source predicted by the present analysis agrees well with the source spectrum preferred by the diffusive reacceleration model. An open question is the significance of the residual path length. This question cannot be fully answered until a new generation of instruments provides measurements of the B/C ratio with greatly improved statistical accuracy in the TeV~amu$^{-1}$ region.

The Leaky-Box model of the present analysis does not include reacceleration in interstellar space as a significant contributor among the effects of Galactic propagation. Consequently, the energy dependence of the propagation path length is fairly strong, with the index $\delta$ close to the often used value of 0.6.

\begin{figure}[tbh]
 \includegraphics[width=\linewidth]{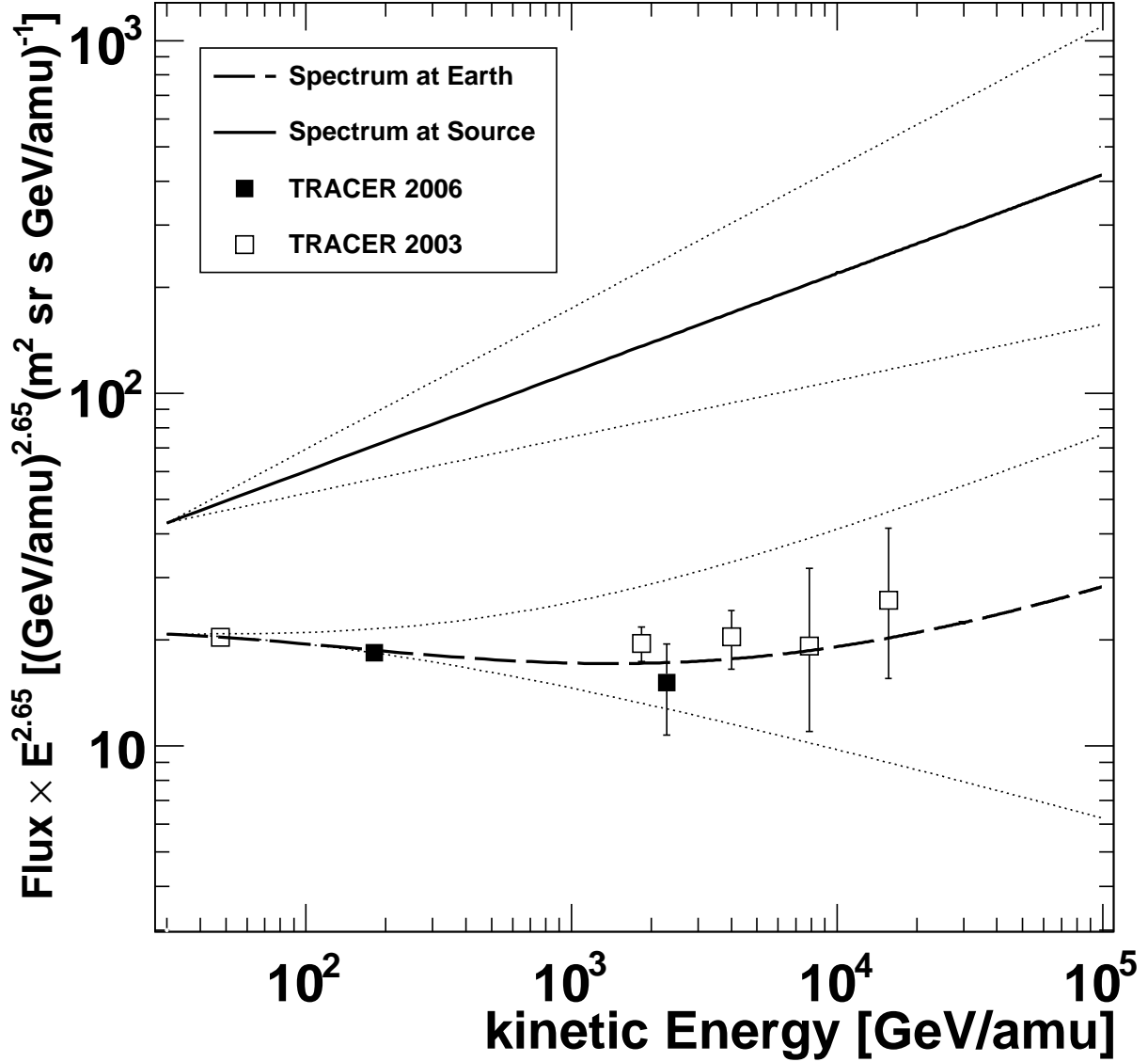}
 \caption{Energy spectrum of oxygen as measured by TRACER. A source spectrum is fit to the data according to Eq.~(\ref{eq:LB_2}) with spectral index $\alpha=2.37\pm0.12$ (solid). The corresponding, observed spectrum is shown as a dashed line and uncertainty bounds are indicated as dotted lines. The lower uncertainty bound corresponds to $\alpha=2.49$, $\delta=0.59$, and $\Lambda_0=0$. The upper bound corresponds to $\alpha=2.25$, $\delta=0.47$, and $\Lambda_0=0.8$~g~cm$^{-2}$. The normalization of the source spectrum is arbitrary. \label{fig6}}
\end{figure}

\begin{figure}[tbh]
 \includegraphics[width=\linewidth]{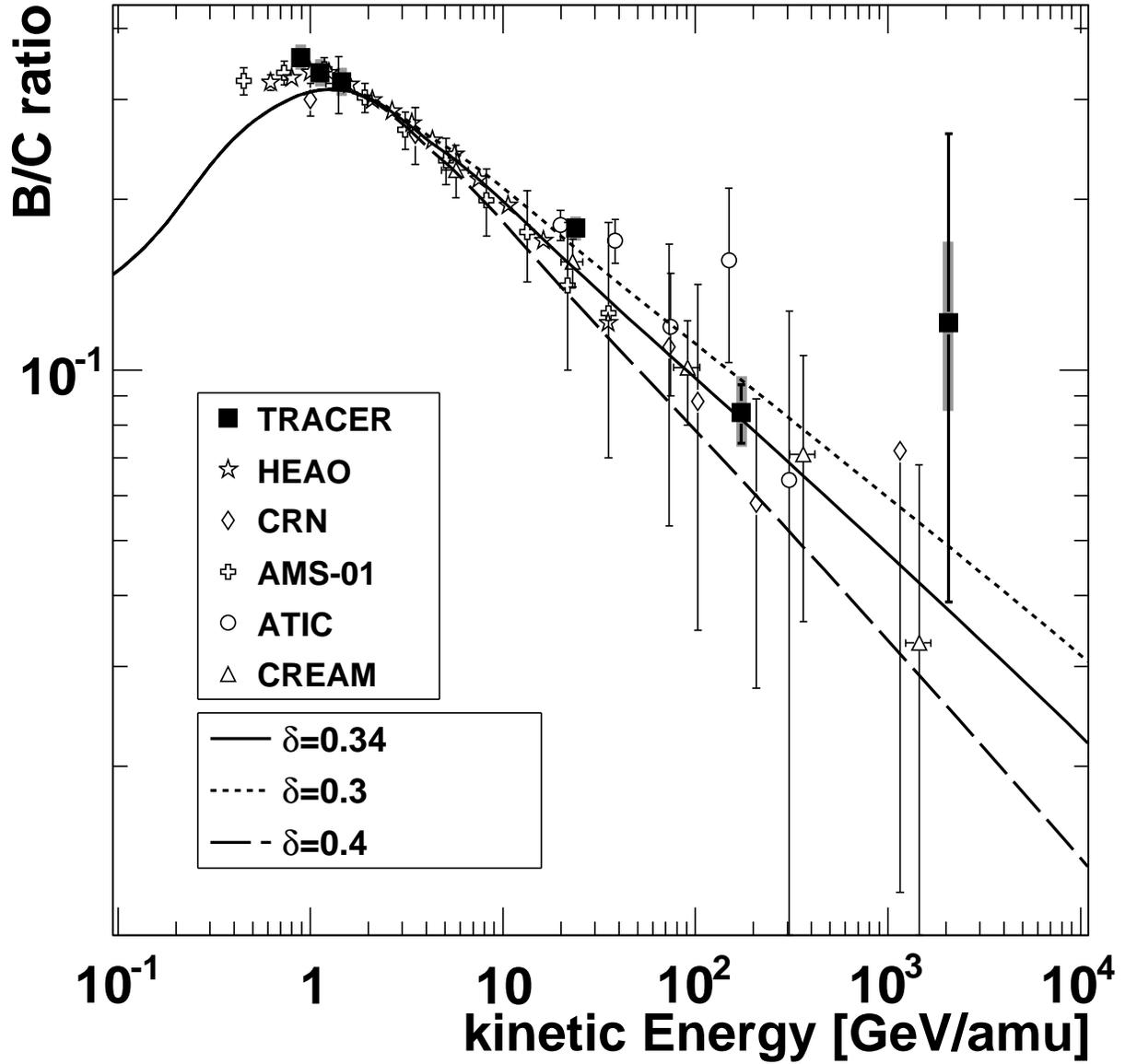}
 \caption{Illustration of diffusive reacceleration models realized with GALPROP for different values for $\delta$. The best model fit to all available cosmic-ray data has $\delta=0.34$ (solid line).\label{fig7}}
\end{figure}

Finally, one may consider deviations from a pure power law behavior of the cosmic-ray energy spectra either at the sources, or at the observation site. In the present analysis, a strict power law was accepted for the source spectrum. As Figure~\ref{fig6} indicates, it would then be inevitable that the observed spectrum deviates from a power law form, but the deviation is so small that it cannot be observed within current observational uncertainties. Or, if detected, it would be difficult to ascribe small effects in the observed spectra to either propagation or source effects.  Once again, much improved statistical and systematic accuracy in future measurements is required to settle this issue.

In summary, it appears that the recent cosmic-ray measurements at high energy are beginning to probe details of the source and propagation characteristics that have remained unexplored for a long time, and one may expect more answers to the remaining questions from a next generation of instruments.

\acknowledgments
We are grateful to M.\ Ave and M.\ Ishimura for helpful discussions and support. AO acknowledges support of FOM in the Netherlands (``Stichting voor Fundamenteel Onderzoek der Materie''). This work was supported by NASA through grants NNG 04WC08G, NNG 06WC05G, and NNX 08AC41G.

\clearpage

\end{document}